\documentclass[11pt]{article}

\usepackage{a4wide}
\usepackage{amsmath}
\usepackage{amsfonts}
\usepackage{amssymb}
\usepackage{euscript}
\usepackage[dvips]{epsfig}
\usepackage{pstricks}

\newtheorem{theorem}{Theorem}
\newtheorem{lemma}[theorem]{Lemma}

\newtheorem{definition}[theorem]{Definition}
\newcommand{\beeq}{\begin{eqnarray*}}
\newcommand{\eneq}{\end{eqnarray*}}

\newenvironment{proof}{\noindent {\it Proof~:}\ }{\ \rule{1mm}{2mm}\medskip}

\newcommand{\N}{\EuScript N}
\newcommand{\Z}{\EuScript Z}
\newcommand{\T}{\EuScript T}
\newcommand{\length}{\EuScript L}
\newcommand{\x}{\mathbf x}
\newcommand{\y}{\mathbf y}
\newcommand{\s}{\mathbf s}
\renewcommand{\u}{\mathbf u}
\renewcommand{\v}{\mathbf v}
\renewcommand{\r}{\mathbf r}

\begin{document}
\title{On quasi-cyclic interleavers for parallel turbo codes
\thanks{Presented in part at
the 2004 IEEE International Symposium on Information Theory, Chicago,
Illinois, USA.}}
\author{Joseph J. Boutros and Gilles Z\'emor
\thanks{J.J. Boutros and G. Z\'emor are with \'Ecole Nationale
Sup\'erieure des T\'el\'ecommunications, 46 rue Barrault, 75634 Paris
13, France. Email: { \{boutros,zemor\}@enst.fr}}}
\date{January 31, 2005}
\maketitle

\begin{abstract}
 We present an interleaving scheme that yields quasi-cyclic turbo
 codes. We prove that randomly chosen members of this family yield
 with probability almost $1$
 turbo codes with asymptotically optimum minimum distance,
 i.e. growing as a logarithm of the interleaver size. These
 interleavers are also very practical in terms of memory requirements and
 their decoding error probabilities for small block lengths compare
 favorably with previous interleaving schemes.
\end{abstract}
\section{Introduction}
It is now well known that the behaviour of turbo codes, although very
powerful under high noise, exhibits an error floor phenomenon that can
be explained by poor minimum distance properties. More specifically,
it can be shown that for randomly chosen interleavers, the expected
minimum distance of a classical two-level turbo code remains constant 
\cite{Perez1996}\cite{Kahale1997}, i.e. does not grow with block length.
Can the error floor behaviour of turbo codes be improved
by designing the interleaver in a way that differs from pure random
choice~? 

This question has been addressed by many authors and the answer
has been shown to be affirmative.
Two types of approaches have been used in trying to find improved
interleavers: the first tries to modify as little as possible a
randomly chosen interleaver by combinatorially avoiding configurations
that yield small-weight codewords. This is in principle
manageable since the expected number of
codewords of constant weight in a turbo code remains constant (does
not grow with block length). Indeed, this approach has met with significant
success:
perhaps the most widely known  design of this type
is the {\em $S$-random} interleaver \cite{Divsalar1995}, that focuses on eliminating
codewords of low weight corresponding
to information sequences of weight $2$. A more recent scheme of
Truhachev et al. \cite{Truhachev2001} weeds out all small-weight turbo-codewords in a
way that is reminiscent of Gallager's method of excluding small cycles
when constructing the parity-check matrix of an LDPC code, see also \cite{Bazzi2003}
for a similar result. They obtain
turbo codes whose minimum distance grows proportionally to $\log N$
where $N$ denotes the interleaver size. This asymptotic result is essentially the best
possible since it was shown by Breiling \cite{Breiling2001} that $D/\log N$
must be upper bounded by a constant, where $D$ and $N$ denote
respectively the minimum distance and interleaver size of the turbo code.

The second approach tries to find interleavers with structure, in
particular algebraic structure, rather than mimic random choice.
Besides enhanced performance, an additional motivation is to have
a permutation with a short description that will save on the memory
required to store the interleaver connections. With this last
feature in mind, a particularly
promising family of interleavers was proposed by Tanner in~\cite{Tanner2000} 
and consists of {\em quasi-cyclic} permutations that
yield quasi-cyclic turbo codes. Encouraging simulation results for the simpler
RA codes were obtained in \cite{Tanner1999}, hinting at good minimum
distance properties of quasi-cyclic turbo-like codes. 
To quote from the conclusion of \cite{Tanner2000}:

{\em ``We conjecture that the class of quasi-cyclic permutations
that will create quasi-cyclic turbo codes is rich, rich enough to
contain codes that will perform as well or better than random
interleavers.''}

In the present work we take up this challenge and study {\em random}
quasi-cyclic permutations. Our approach borrows from both the
unstructured, almost random, and the algebraic, structured, design
strategies: our interleavers have structure, and the inherent
advantageous storage properties, and yet involve a certain amount of
random choice. Our main result is to show that the typical minimum
distance of the associated turbo code grows linearly with $\log N$,
i.e. has optimal growth, thus justifying Tanner's conjecture. 
Furthermore, for moderate lengths these interleavers turn
out to be not only practical, but very efficient, 
comparing favorably with random, 
$S$-random, and all known interleavers in a number of instances.

The paper is organized as follows. In section 2 we give a short
summary of previous work on interleaver design. In section 3 we
describe our family of quasi-cyclic interleavers. Section 4 is devoted
to proving that a randomly chosen interleaver from this family will
have asymptotically optimal minimum distance with high probability 
(Theorem~\ref{th:main}).
Section 5 gives experimental results for short lengths ($N=400$ and $N=1600$).

\section{Previous work on interleaver design}
Classical channel coding systems using a serial concatenation of Reed-Solomon codes 
and binary convolutional codes include a matrix interleaver (also called {\em block interleaver}) 
which enables one to split the error bursts generated by the Viterbi decoder before applying
an algebraic Berlekamp-Massey decoding \cite{Blahut2003}\cite{Lin2004}.
The early research on interleavers for digital communications has been preceded
by the invention of burst-error-correcting cyclic and burst-error-correcting 
convolutional codes \cite{Peterson1972}.
Although the low density parity check codes developed by R. Gallager 
\cite{Gallager1963} integrated random interleaving of the parity-check
matrix columns, no serious study on interleavers was known until the work by Ramsey
on optimum interleavers for infinite length sequences \cite{Ramsey1970}. 
For example, a type I Ramsey interleaver 
guarantees an input separation $n_1$ and an output separation $n_2$ with a minimum
delay equal to $n_2(n_1-1)$, where $n_1$ and $n_2$ are two positive integers
satisfying $n_2 < n_1 < 2n_2$, $n_1$ and $n_2+1$ are relatively prime.

The separation guaranteed by Ramsey interleavers has
been named {\em spreading} after the invention of parallel turbo codes
based on binary systematic recursive convolutional constituents \cite{Berrou1993}\cite{Berrou1996}.
Finite length interleavers or permutations designed for parallel turbo codes
have been extensively studied during the last decade. 
The amount of publications on the subject ranges in the hundreds and
cannot be listed here in full. The following selection of interleaver
families is an attempt to give a meaningful picture of the state of
research and to summarize the main techniques.
As mentioned
earlier they can be crudely partitioned into two categories:
mostly random interleavers with a weak structure, requiring
an exhaustive description of the permutation, or strongly structured
with short representations.

\begin{itemize}
\item {\bf Purely random interleavers}.

These interleavers are built from permutations on $N$ integers selected at random.
Here, $N$ denotes the interleaver size. The original turbo codes ($N=65536$ bits)
were designed with purely random interleavers. Without any interleaver optimization,
the error rate performance of parallel turbo codes can be enhanced via primitive
feedback polynomials in the turbo code convolutional constituents \cite{Benedetto1996}.
\item {\bf Random interleavers with a weak deterministic structure}.

This family includes the S-random or spread interleaver proposed by Divsalar and Pollara \cite{Divsalar1995}. 
The S-random interleaver $\pi$ is constructed at random, it must satisfy
the constraint $|\pi(i)-\pi(j)|>S$ for all $|i-j|<S$, where the maximal theoretical
value of the spread $S$ is $\sqrt{N}$. High spread random (HSR) interleavers proposed 
by Crozier \cite{Crozier2000} belong to this family. They rely on the maximization of
the spread $S=\min \{ |i-j|+|\pi(i)-\pi(j)| \}$ (also defined in \cite{Andrews1998} for arithmetic
and random interleavers). The spread of HSR interleavers is upper bounded by $\sqrt{2N}$.
The permutation described by Truhachev et.al. \cite{Truhachev2001} that guarantees an asymptotically
optimal minimum distance is mostly random with a weak deterministic structure.
\item {\bf Deterministic algebraic/arithmetic interleavers}.

Many algebraic permutations have been suggested or specifically developed for parallel turbo codes.
In spite of (or perhaps because of) their very low memory, they tend to
exhibit intermediate or poor error rate performance.
A short selection consists of
the interleavers described by Berrou and Glavieux \cite{Berrou1996},
by Andrews et.al. \cite{Andrews1998}, Sadjadpour et.al. \cite{Sadjadpour2001}, 
Bravo and Kumar \cite{Bravo2003}. The Relative Prime and the Golden interleavers
described by Crozier et.al. \cite{Crozier1999} belong to this family 
of deterministic interleavers.
\item {\bf Deterministic interleavers with a weak random structure}.

We mention two types of deterministic interleavers where randomness has
been added in order to unbalance somewhat the algebraic structure. 
Dithered golden interleavers \cite{Crozier1999}
and dithered relative prime (DRP) interleavers \cite{Crozier2001}.
DRP interleavers exhibit excellent error rate performance.
They are obtained in 3 steps: 1- application of a small permutation (input dithering) 
to the interleaver input, e.g., a size 8 permutation applied $N/8$ times,
2- a relative prime permutation $j=s+ip$, where $j$ is the read position, $i$ is the write position,
$s$ is a shift and $p$ is prime relative to $N$, 3- an output dithering similar to the input one.
\item {\bf Interleavers from the graphical structure of codes}.

Cayley-Katz graphs with large girth have been used to design Generalized Low Density (GLD)
codes with binary BCH constituents \cite{Pothier2000}. Similar application
was made by Vontobel \cite{Vontobel2002} to design turbo code interleavers
from large girth graphs. Interleavers based on large girth graphs are all deterministic.
Yu et.al. \cite{Yu2002} also designed good interleavers by looking at the 
loop distribution in the turbo code structure. Such interleavers are random
with a weak deterministic structure.
\item {\bf Interleavers by other criteria}.

Abbasfar and Yao \cite{Abbasfar2004} recently proposed
good interleavers that eliminate codewords with Hamming weight 
less than a certain distance. The construction algorithm is based on a two dimensional 
representation of the permutation. This representation previously inspired Crozier \cite{Crozier2000}
in his design of dithered-diagonal interleavers. The interleaver 
design by distance spectrum shaping can be classified in 
the class of random interleavers with a weak deterministic structure.
Finally, we mention the interleavers designed by Hokfelt et.al. \cite{Hokfelt1999} 
via the minimization of the correlation between extrinsic informations under iterative decoding. 
\end{itemize}

The bi-dimensional quasi-cyclic interleaver described in the next section 
combines randomness and determinism in an almost equal manner.
When designed from a square matrix, 
its quasi-cyclicity period is $\sqrt{N}$, meaning that a set of $2\sqrt{N}$
integers is needed to save the bi-dimensional interleaver into memory, 
rather than the $N$ integers needed for a purely random permutation.

\section{Bi-dimensional, or quasi-cyclic interleavers}\label{sec:permutation}
For simplicity, we restrict ourselves to the classical turbo code construction
with a fixed constituent convolutional code $C_0$ of rate $R_0=1/2$. 
The turbo
encoder takes an information sequence $\s$ of $N$ bits, produces a first 
sequence of $N$ check bits by submitting $\s =s_0,s_1,\ldots s_{N-1}$ to 
an encoder for $C_0$,
and a second sequence of a further $N$ check bits by submitting
a permuted version $s_{\pi(0)},s_{\pi(1)},\ldots ,s_{\pi(N-1)},$
of $\s$ to the encoder for $C_0$. The overall turbo code rate is $R=1/3$
and the {\em interleaver} is the permutation $\pi$ on the ordered set
of information coordinates $\N =\{0,1,\ldots ,N-1\}$.

There will be a $2$-dimensional structure inherent to our
choice of permutation $\pi$,
therefore we shall restrict ourselves to the case when 
$N=n_1\times n_2$ is a composite integer.

Let $\pi$ be a permutation on $\N =\{0,1,\ldots ,N-1\}$
defined as follows. For any 
$(i,j)\in\{0,1,\ldots,n_1-1\}\times\{0,1,\ldots,n_2-1\}$ define
the function
\beeq
\phi : \{0,1,\ldots,n_1-1\}\times\{0,1,\ldots,n_2-1\} & \rightarrow & \N\\
 (i,j)               & \mapsto     & i\times n_2+j
\eneq
Let $\sigma$ be a permutation of $\{0,1,\ldots ,n_2-1\}$ and let
$(X_j)_{j=0\ldots n_2-1}$ be a family of integers mod $n_1$. Define
the permutation $\Pi$ on $\{0,1,\ldots,n_1-1\}\times\{0,1,\ldots,n_2-1\}$ by
  $$\Pi(i,j) = (i+X_j \bmod n_1, \sigma(j)).$$
Finally define the permutation $\pi = \phi\Pi\phi^{-1}$
on the set $\N$. A small example is given in Figure~\ref{fig:pi}.

\begin{figure}
  \centering
 $\sigma = \begin{pmatrix}
0 & 1 & 2 & 3 & 4\\ 3 & 2 &0 & 4 & 1\end{pmatrix}\hspace{1cm}
X_0=0, X_1=3, X_2=4, X_3=2, X_4=1.$
$$\begin{array}{ccc}
\begin{bmatrix}
    0 & 1 & 2 & 3 & 4\\
    5 & 6 & 7 & 8 & 9\\
    10 & 11 & 12 & 13 &14\\
    15 & 16 & 17 & 18 & 19\\
    20 & 21 & 22 & 23 & 24
  \end{bmatrix}
 &
  \begin{bmatrix}
    3 & 2 & 0 & 4 & 1\\
    8 & 7 & 5 & 9 & 6\\
    13 & 12 & 10 & 14 &11\\
    18 & 17 & 15 & 19 & 16\\
    23 & 22 & 20 & 24 & 21
  \end{bmatrix}
&
    \begin{bmatrix}
    3 & 12 & 5 & 19 & 21\\
    8 & 17 & 10 & 24 & 1\\
    13 & 22 & 15 & 4 & 6\\
    18 & 2 & 20 & 9 & 11\\
    23 & 7 & 0 & 14 & 16
\end{bmatrix}\\[2mm]
{\mathbf A} & {\mathbf B} & {\mathbf C}
\end{array}$$
$$
\begin{array}{c}
(\pi(0),\pi(1),\pi(2),\ldots ,\pi(24))=\\
(3,12,5,19,21,8,17,10,24,1,13,22,15,4,6,18,2,20,9,11,23,7,0,14,16).
\end{array}
$$
  \caption{Example: construction of $\pi$, $N=25$, $n_1=n_2=5.$ 
Write $0,1,\ldots , N-1$ in a square array ${\mathbf A}$, apply
$\sigma$ to permute columns, giving ${\mathbf B}$, and
rotate column $j$ cyclically, by $X_j$ $\bmod\; 5$, giving array
${\mathbf C}$. Read off the rows to get $\pi(0),\pi(1),\pi(2),\ldots$.
}
  \label{fig:pi}
\end{figure}

The quasi-cyclic nature of the permutation $\pi$ just defined is
stressed in the following Lemma, a direct consequence of the
definition.

\begin{lemma}\label{lem:quasicyclic}
  A permutation $\pi$ belonging to the class defined above satisfies,
  for any $x,x'\in \N$ such that $x'=x+n_2 \bmod N$,
    $$\pi(x')=\pi(x)+n_2  \bmod N.$$
\end{lemma}

If we make the trellis of the constituent convolutional code
tail-biting, and if we write the check bits of the turbo code in the
proper order, we obtain a quasi-cyclic turbo code \cite{Tanner2000}. For
this reason, permutations of the above type will be called
{\em $(n_1,n_2)$-quasi-cyclic} (or simply quasi-cyclic).

We shall take instances of $(n_1,n_2)$-quasi-cyclic permutations $\pi$
by choosing the permutation $\sigma$ randomly, with uniform distribution,
among permutations of $\{0,1,\ldots,n_2-1\}$, and by choosing
the $X_i,i=0\ldots n_1-1$ randomly, with uniform distribution, among
the set of integers mod $n_1$. We choose the $X_i$ to be independent of
each other and of $\sigma$. This is a way of choosing $\pi$ uniformly
in the class of $(n_1,n_2)$-quasi-cyclic permutations.

As a first comment, we may note that $\pi$ has quite a lot more
structure than a totally random permutation. A certain amount of
randomness remains however; to quantify it somewhat, suppose for
example that $n_1=n_2=n=\sqrt{N}$ (we shall see experimentally in section \ref{sec_simuls}
that $n_1=n_2$ is a good choice), we see that
$\pi$ is defined by $\log n! +n\log n \approx \sqrt{N}\log N$ 
random  bits as opposed to the $N\log N$ bits that define an otherwise 
unstructured permutation.

Our strategy will be probabilistic, i.e. we will estimate the probability
that the permutation $\pi$ produces turbo code words of small weight
$w << \log N$ and show that this probability must be vanishingly small.
Interestingly, over all permutations $\pi$, the {\em expected} number of
turbo codewords of small weight $w$ does not vanish with $N$. This is
because of an all-or-nothing phenomenon. Permutations of the above type
produce either no turbo codewords of small weight, or relatively many
(at least $n=\sqrt{N}$).

\section{Minimum distance analysis}

For any two integers $x$ and $y$ of $\N$ let us denote by $d(x,y)$
the circular distance between $x$ and $y$, i.e.
the smallest non-negative integer $d$ such that $x+d=y \bmod N$
or $x-d = y \bmod N$. Let us draw an edge between $x$ and $y$ whenever
$d(x,y)=1$, giving $\N$ a circular structure: 
by an interval of $\N$ we shall mean a sub-path of $\N$.

Let $\s=s_0,\ldots ,s_{N-1}$
be an information sequence, and let $\v\subset\N$ be the support of~$\s$.
The information sequence $\s$ generates a path in the (tail-biting) 
trellis of the convolutional code $C_0$.
Consider the partition $\N=\Z\cup \T$ where $\Z$ is defined as the set of
coordinates $i$ for which the path associated to $\s$ goes from the zero
state to the zero state. For every $i\in\Z$ we have $s_i=0$ and the
corresponding check bit is also $0$. The complement $\T$ of $\Z$ is
a union of intervals 
$\T=[a_1,b_1]\cup [a_2,b_2]\cup\ldots \cup [a_m,b_m]$.
The intervals $[a_j,b_j]$ 
are sometimes called simple trellis paths, or simple error events in the convolutional coding terminology.
The trellis of a recursive convolutional code has the property that
the zero state can only be left at time $t$ if $s_t=1$ and it can only
be reached from a nonzero state at time $t$ if $s_t=1$. This means
that $\v\subset \T$ and every interval $[a_j,b_j],j=1\ldots m$
starts and ends with an element of the support $\v$ of $\s$.
A recursive convolutional code also has the property of outputting a
steady stream of non-zero symbols during the time it goes
through a simple trellis path, i.e. during the time it is fed the
information bits $s_t$ for $t\in [a_j,b_j]$. In other words, there exists
a constant $\lambda$, depending only on $C_0$, such that the total weight
of the convolutional codeword associated to the information sequence $\s$ is
at least $\lambda\sum_{j=1}^m d(a_j,b_j)$. Some numerical values
of $\lambda$ are given in Table \ref{table:lambda} by way of
illustration. Summarizing:

\medskip

\noindent
{\bf Facts:} Associated to any information sequence $\s$ of support
$\v$ there is a subset $\T(\s)\subset \N$ such that
\begin{enumerate}
\item $\T(\s)=[a_1,b_1]\cup [a_2,b_2]\cup\ldots \cup
[a_m,b_m]$ is a union of intervals of $\N$
\item $\v\subset\T(\s)$
\item for any $j=1\ldots m$, $|\v\cap [a_j,b_j]|\geq 2$
\item on input $\s$ the convolutional encoder outputs at least 
      $\lambda\sum_{j=1}^md(a_j,b_j)$ nonzero symbols,
      for some positive constant $\lambda$
\end{enumerate}

\begin{table}[t]
{\small
\begin{center}
\begin{tabular}{|c|c|c|}
\hline
\multicolumn{3}{|c|}{ \textbf{Rate 1/2 RSC codes}}\\\hline \hline
Octal Generators  &  Number of States & Parameter $\lambda$ \\ \hline
(7, 5)   &  4  & 1/2  \\ \hline
(13, 15) &  8  & 2/5  \\ \hline
(17, 15) &  8  & 1/2  \\ \hline
(37, 21) &  16 & 1/4  \\ \hline
(23, 35) &  16 & ~4/11 \\ \hline
\end{tabular}
\end{center}
}
\caption{Rate 1/2 recursive systematic convolutional codes. The parameter $\lambda$ is the minimal ratio 
of Hamming weight to trellis length among all codewords.}\label{table:lambda}
\label{TableLambda}
\end{table}

Let us call the {\em trellis weight} of $\s$ the quantity 
$\sum_{j=1}^md(a_j,b_j)$ defined above, denote it by $W_T(\s)$.
Now the turbo code word associated to $\s$ has its weight lower-bounded
by both convolutional codewords corresponding to the input $\s$ and
to the permuted input $\s^\pi$. Since the maximum is lower-bounded by
half the sum, Fact 4 above implies:

\begin{lemma}\label{lem:trellisweight}
  If the information sequence $\s$ produces a turbo codeword of
  Hamming weight $w$, then $W_T(\s)+W_T(\s^\pi)\leq 2w/\lambda$.
\end{lemma}

This last lemma says that low-weight turbo codewords can only exist if
there is an information sequence $\s$ such that both $\s$ and $\s^\pi$
have small trellis weight. Now the decomposition of the supports of
$\s$ and $\s^\pi$ into simple trellis paths is rather awkward to
handle probabilistically, so we shall introduce a related concept that
will be easier  to deal with. The following definition is purely
combinatorial.

\begin{definition}
  Let $\x=x_0,x_1,\ldots x_\ell$, $\ell$ odd, be an even-numbered sequence of
  elements of $\N$. Let $y_i=\pi(x_i), i=0\ldots \ell$ for some
  permutation $\pi$ of $\N$. Let us call the {\em $\pi$-weight} of
  $\x$ the quantity:
     $$w_\pi(\x)=\sum_{1\leq i, 2i<\ell}d(x_{2i-1},x_{2i})+d(x_0,x_{\ell}) + 
       \sum_{0\leq i, 2i+1\leq\ell}d(y_{2i},y_{2i+1}).$$
\end{definition}

The reason for introducing the above definition lies in the following
lemma.

\begin{lemma}\label{lem:piweight}
  If there exists a codeword of weight $w$ in the turbo code with
  interleaver $\pi$, then there exists an
  even-numbered sequence $\x$ of distinct elements of $\N$ of
  $\pi$-weight $w_\pi(x) \leq 2w/\lambda$.
\end{lemma}

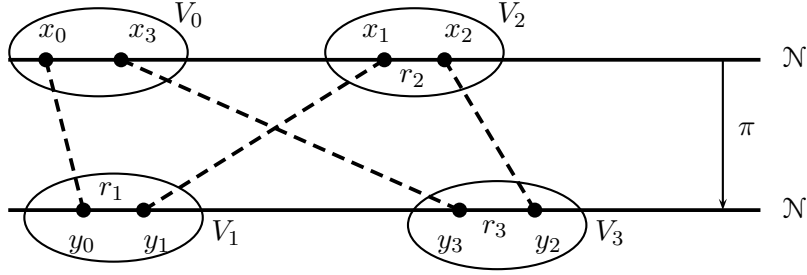
\begin{figure}
\begin{center}
 \begin{pspicture}(11,4)
\psline[linewidth=0.05](0,1)(10,1)
\psline[linewidth=0.05](0,3)(10,3)
\pscircle*(0.5,3){0.1}
\pscircle*(5,3){0.1}
\pscircle*(5.8,3){0.1}
\pscircle*(1.5,3){0.1}
\pscircle*(1,1){0.1}
\pscircle*(1.8,1){0.1}
\pscircle*(6,1){0.1}
\pscircle*(7,1){0.1}
\psline[linewidth=0.05,linestyle=dashed](0.5,3)(1,1)
\psline[linewidth=0.05,linestyle=dashed](1.8,1)(5,3)
\psline[linewidth=0.05,linestyle=dashed](5.8,3)(7,1)
\psline[linewidth=0.05,linestyle=dashed](1.5,3)(6,1)
\psline{->}(9.5,3)(9.5,1)
\put(9.7,2){$\pi$}
\put(0.8,0.5){$y_0$}\put(1.8,0.5){$y_1$}
\put(0.4,3.3){$x_0$}\put(4.7,3.3){$x_1$}
\put(5.8,3.3){$x_2$}\put(1.6,3.3){$x_3$}
\put(5.7,0.5){$y_3$}\put(7,0.5){$y_2$}
\put(1.2,1.2){$r_1$}\put(5.2,2.7){$r_2$}
\put(6.3,0.7){$r_3$}
\psellipse(1.2,3.1)(1.2,0.6)\psellipse(5.4,3.1)(1.2,0.6)
\psellipse(1.4,0.9)(1.2,0.6)\psellipse(6.5,0.8)(1.2,0.6)
\put(2.2,3.5){$V_0$}\put(6.5,3.5){$V_2$}
\put(2.7,0.6){$V_1$}\put(7.8,0.6){$V_3$}
\put(10.3, 2.9){$\N$}\put(10.3, 0.9){$\N$}
\end{pspicture}
\end{center}
\caption{$\ell =3$. The cycle $V_0,V_1,V_2,V_3$ defined in the proof
  of Lemma \ref{lem:piweight}, the associated sequences
  $\x=x_0,x_1,x_2,x_3$, $\y=y_0,y_1,y_2,y_3$, $\r=r_1,r_2,r_3$.}
\label{fig:bipartite}
\end{figure}

\begin{proof}
  Let $\s$ be the information sequence corresponding to the turbo
  codeword of weight $w$ and let $\v$  be its support. Note that
  the support of $\s^\pi$ is $\u =\pi^{-1}(\v)$. 
  Let $\T(\s)=[a_1,b_1]\cup [a_2,b_2]\cup\ldots \cup [a_m,b_m]$ be the
  decomposition of $\T(\s)$ into $m$ intervals given by Fact 1 and let
  $\T(\s^\pi)=[a_1',b_1']\cup [a_2',b_2']\cup\ldots \cup [a_k',b_k']$ be
  the corresponding decomposition for the permuted version $\s^\pi$ of
  the information sequence. Now consider the bipartite graph whose
  vertex set is made up of the two sets $A$ and $B$ where
  $A$ is the set of the $k$ intervals $[a_i',b_i'],j=1\ldots k,$
  and $B$ is the set of the $m$ intervals $[a_j,b_j],j=1\ldots m$.
  Put an edge between interval $[a_i',b_i']$ and
  $[a_j,b_j]$ for every $x\in\pi^{-1}(\v)\cap [a_i',b_i']$ such
  that $\pi(x)\in [a_j,b_j]$ (multiple edges may occur). 
  Fact 2 implies that the minimum degree
  of the bipartite graph is at least $2$. Therefore there exists an
  (even-length) elementary cycle in the graph, i.e. a string of
  distinct vertices $V_0,V_1,\ldots V_{\ell}$, $\ell$ odd, where the interval
  $V_i$ belongs to $A$
  (respectively $B$) for $i$ even (respectively odd) and where there
  is an edge between $V_i$ and $V_j$ whenever $i-j=\pm 1 \bmod
  \ell +1$. For $0\leq i, 2i<\ell$ the edge between $V_{2i}$ and $V_{2i+1}$ is
  defined by an element of $V_{2i}\cap\u$ that we denote
  $x_{2i}$, and an element of $V_{2i+1}\cap\v$ that we denote $y_{2i}$ and
  that equals $y_{2i}=\pi(x_{2i})$. Similarly, for $1\leq i=1,2i<\ell$,
  every edge between $V_{2i}$ and $V_{2i-1}$ is associated to
  $x_{2i-1}\in V_{2i}$ and $y_{2i-1}\in V_{2i-1}$ with 
  $y_{2i-1}=\pi(x_{2i-1})$. Finally let $x_{\ell}\in V_0$ and 
  $y_{\ell}=\pi(x_{\ell})\in V_{\ell}$ correspond to the edge between
  $V_0$ and $V_{\ell}$.
  
  We have constructed a sequence $\x=x_0,x_1,\ldots ,x_{\ell}$ of
  elements of the support $\u =\pi^{-1}(\v)$ of $\s^\pi$  such  that
  $\{x_{2i-1},x_{2i}\}\subset V_{2i}$, $1\leq i, 2i\leq\ell$, 
  $\{x_{\ell},x_{0}\}\subset V_0$, and $\{y_{2i},y_{2i+1}\}\subset
  V_{2i+1}$, $0\leq i, 2i+1\leq\ell$: see Figure \ref{fig:bipartite}. 
  Therefore, denoting by $\length(V)$ the
  length of an interval $V$, we have
  $$w_\pi(\x) \leq \sum_{i=0}^{\ell}\length(V_i)\leq W_T(\s)+W_T(s^\pi)$$
which proves the result by Lemma \ref{lem:trellisweight}.
\end{proof}

We shall now study the probability that a sequence of small
$\pi$-weight exists. We need some more notation.

  Let $\r=r_1,r_2,\ldots ,r_\ell$, $\ell$ odd, be a sequence
 of integers modulo $N$. Let $|r_i|$ denote the smallest absolute value of
 a (possibly negative) integer equal to $r_i$ modulo $N$, and let
 $\|\r\|=|r_1|+|r_2|+\ldots +|r_\ell|$.
Let $x_0\in \N$.
Together $x_0$ and $\r$ uniquely define the $(\ell +1)$-sequence
 $\x= \x(\r,x_0) = x_0,x_1,\ldots ,x_\ell$
and $\y =\y(\r,x_0)= y_0,\ldots ,y_\ell$ such that
$$\begin{array}{llrrcl}
1. & y_i = \pi(x_i), i=0\ldots\ell, &&&&\\
2. & \text{for all $i\geq 0$ such that} & 2i+1\leq \ell, &
      y_{2i+1} & = & y_{2i}+r_{2i+1} \bmod N\\
3. & \text{for all $i\geq 1$ such that} & 2i<\ell, &
      x_{2i} & = & x_{2i-1}+r_{2i} \bmod N
\end{array}$$

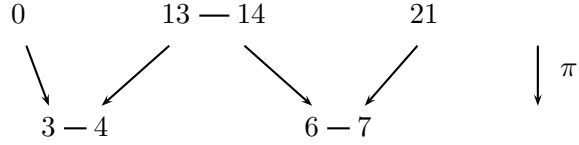
\begin{figure}
  \centering
  \begin{pspicture}(8,2)
    \put(0,1.5){$0$}\psline{->}(0.2,1.2)(0.5,0.4)\put(0.4,0){$3$}
    \psline(0.7,0.1)(1,0.1)\put(1.1,0){$4$}
    \psline{->}(2.1,1.2)(1.2,0.4)\put(2,1.5){$13$}
    \psline(2.5,1.6)(2.9,1.6)\put(3,1.5){$14$}
    \psline{->}(3.1,1.2)(4,0.4)\put(3.9,0){$6$}
    \psline(4.2,0.1)(4.5,0.1)\put(4.6,0){$7$}
    \psline{->}(5.4,1.2)(4.7,0.4)\put(5.3,1.5){$21$}
    \psline{->}(7,1.2)(7,0.4)\put(7.3,0.8){$\pi$}
  \end{pspicture}
  \caption{Let $\pi$ be the same as in Figure
    \ref{fig:pi}. The sequence $\r=1,1,1$ together with $x_0=0$
    define $\x=\x(\r,x_0)=0,13,14,21$.
   We have $w_\pi(\x) = 1+1+1+d(0,21)=7$.}
  \label{fig:Mcycle}
\end{figure}

Note that 
\begin{equation}
  \label{eq:r,x_0}
  w_\pi(\x(\r,x_0)) =\|\r\| + d(x_0,x_\ell).
\end{equation}
Finally, let us say that the sequence $\r$ {\em $M$-cycles} at $x_0$ if
  $$w_\pi(\x(\r,x_0)) \leq M.$$
The definitions are illustrated in Figure \ref{fig:Mcycle}.

Lemma \ref{lem:piweight} translates directly into the following, given
that the $x_i$ are distinct  only if all the $r_i$ are non-zero.

\begin{lemma}\label{lemma:scatters}
  If there exists a turbo codeword of weight $w$ then there exists
  $x_0\in\N$ and a non-zero sequence $\r=r_1,\ldots ,r_\ell$, 
  $r_i\neq 0,i=1\ldots\ell$, $\ell$ is odd, that $2w/\lambda$-cycles at $x_0$.
\end{lemma}

We now have everything in place for doing the probabilistic analysis.
Let $Z_{\r,x_0}$ be the Bernoulli random variable equal to $1$ if 
the sequence $\r$
$M$-cycles at $x_0$ and equal to $0$ otherwise. The set of all
permutations $\pi$ of the set $\N$ is endowed with two probability
measures, namely:
\begin{itemize}
\item the uniform probability measure $P_r$, in other words
  $P_r(\pi)=1/N!$ for all $\pi$.
\item the quasi-cyclic probability measure $P_q$ defined
  by $P_q(\pi)=1/(n_1^{n_2}n_2!)$ if $\pi$ is $(n_1,n_2)$-quasi-cyclic 
  and $P_q(\pi)=0$ otherwise. Note that, as mentioned in section 
\ref{sec:permutation}, this is equivalent to choosing 
the permutation $\sigma$ randomly, with uniform distribution,
among permutations of $\{0,1,\ldots,n_2-1\}$, and by choosing
the $X_i,i=0\ldots n_1-1$ randomly, independently of each other and of
$\sigma$, and with uniform distribution, among
the set of integers mod $n_1$. 
\end{itemize}

\begin{lemma}\label{lem:Zrx_0}
Let $M<N$, $x_0$ and $\r=r_1,\ldots ,r_\ell$ be given, $r_i\neq 0,
i=1\ldots\ell$, $\ell$ is odd. 
  We have:
  \begin{enumerate}
  \item If $\|\r\|\geq M$ then $P_r[Z_{\r,x_0}=1]=P_q[Z_{\r,x_0}=1]=0$.
  \item If $\|\r\| < M$  then $P_r[Z_{\r,x_0}=1]\leq 2M/(N-1)$.
  \item If $\|\r\| < M <n_2$, then 
        $P_q[Z_{\r,x_0}=1]\leq \frac{2M}{n_1(n_2-1)}$.
  \item If $\|\r\| < M$ and $r_i=0 \bmod n_2$ for every $i=1\ldots
    \ell$, then $P_q[Z_{\r,x_0}=1]=1$.
  \end{enumerate}
\end{lemma}

\begin{proof}
  Point 1 is a direct consequence of \eqref{eq:r,x_0}.

  To see Point 2, consider $x_1,\ldots , x_\ell$ as random variables.
  Conditional on the position of $x_{\ell -1}$, i.e. on the event
  $x_{\ell -1}=k$, the distribution of $x_\ell$ is, since 
  $r_{\ell}\neq 0$, uniform on the set $\N\setminus\{k\}$.
  Therefore
  \beeq
  P_r(Z_{\r,x_0}=1) &=&
  \sum_kP_r[d(x_\ell,x_0)\leq M-\|\r\|\;|\; x_{\ell -1}=k]
  P_r[x_{\ell -1}=k]\\
  &\leq & \sum_k\frac{1+2(M-\|\r\|)}{N-1}P_r[x_{\ell -1}=k] 
  ~~=~~\frac{1+2(M-\|\r\|)}{N-1}\\
  &\leq &\frac{2M}{N-1}.
  \eneq
  
  To see Point 3 argue as follows: write $d(x_0,x_\ell) = qn_2+\rho$,
  $0\leq\rho<n_2$. Since $M<n_2$ we have $Z_{\r,x_0}=1$ if and only if 
  \begin{itemize}
  \item[$(a)$] $q=0$
  \item[$(b)$] $\rho\leq M-\|\r\|$.
  \end{itemize}
  Since $r_{\ell}\neq 0$ and $M<n_2$ imply that $r_{\ell}\neq 0 \bmod
  n_2$, we can argue as in point 3, replacing the random permutation
  $\pi$ of $\{0,1,\ldots ,N-1\}$ by the random permutation $\sigma$ of
  $\{0,1,\ldots ,n_2-1\}$, to obtain that the event $(b)$ occurs with
  probability not more than $2M/(n_2-1)$. 
  By construction of the quasi-cyclic permutation the event $(a)$ is
  independent of $(b)$ and occurs with probability $1/n_1$. 

  Point $4$ is simply due to the fact that for quasi-cyclic $\pi$ we
  have $\pi(x_0+n_2)=\pi(x_0)+n_2 \bmod N$ for any
  $x_0\in \N$, therefore
  $d(x_0,x_\ell)=\sum_{1\leq i\leq\ell}r_i < M$.
\end{proof}

Next, we shall study the expected number of couples $(\r,x_0)$ for which
$\r$ $M$-cycles  at $x_0$, i.e. the expectation of the random
variable
\begin{equation}
  \label{eq:Z}
  Z = \sum_{x_0\in\N, r_i\neq 0, \|\r\|<M}Z_{\r,x_0}.
\end{equation}
Since $\r$ must have only non-zero terms, its length $\ell$ cannot
exceed its norm $\|\r\|$. The number of sequences of given norm
$m$, length $\ell$ and non-negative terms is exactly 
$\binom{m}{\ell}$, so that the number of terms in the
sum \eqref{eq:Z} is not more than
  $$N\sum_{1\leq m< M}\sum_{0\leq\ell\leq m}2^\ell\binom{m}{\ell}=
     N\sum_{1\leq m< M}3^m\leq N3^M/2.$$
{From} this and Point 3 of Lemma \ref{lem:Zrx_0} we obtain therefore:
\begin{lemma}\label{lem:expected}
  Let $M<n_2$. The expected number $E_q[Z]$ of couples $(\r,x_0)$ such
  that $\r$ $M$-cycles at $x_0$ satisfies, for the probability measure
  $P_q$,
    $$E_q[Z] \leq M3^{M}(1-1/n_2)^{-1}.$$
\end{lemma}

Notice that Point 2 of Lemma \ref{lem:Zrx_0} would give essentially the
same estimate of the expected value of $Z$ for uniformly random $\pi$.
However, the crucial property of the class of quasi-cyclic permutations
that will make a big difference between choosing $\pi$ uniformly
random and quasi-cyclic-random is the following direct consequence of
Lemma \ref{lem:quasicyclic}:

\begin{lemma}
  If the sequence $\r$  $M$-cycles at $x_0$ for a quasi-cyclic $\pi$, then $\r$
  $M$-cycles at $x_0+n_2\bmod N$ for $\pi$. In particular, $Z$ is a
  multiple of $n_1$ for every quasi-cyclic permutation $\pi$.
\end{lemma}

This means that $E_q[Z]=\sum_{z\geq n_1}zP_q[Z=z]\geq n_1P_q[Z>0]$.
We have therefore that whenever the quantity $E_q[Z]/n_1$ is made to
be vanishing with $N$, the probability that there exists a sequence
$\x$ of $\pi$-weight not more than $M$ tends to zero. Putting together
Lemma \ref{lem:expected} and Lemma \ref{lem:piweight} we obtain this
section's main result~:

\begin{theorem}\label{th:main}
For any constant $C< \lambda/2$ and block length $N=n_1n_2$ chosen to
satisfy $n_2>\frac{2C}{\lambda}\log_3n_1$,
the minimum distance of the random quasi-cyclic turbo code satisfies
  $$D\geq C\log_3 n_1$$
with probability that tends to $1$ as $n_1$ tends to infinity.
\end{theorem}

\section{Experimental results for practical lengths \label{sec_simuls}}
In this section, we provide computer simulation results for the word error rate (WER) 
of parallel turbo codes using our new family of interleavers 
and comparing it with $S$-random and random interleavers.
The output of the turbo encoder is modulated via a binary phase shift keying (BPSK)
modulation and transmitted over an ideal additive white gaussian noise (AWGN) channel.
The turbo decoder performs iterative a posteriori probability estimation 
by applying the forward-backward algorithm to each convolutional constituent.

\begin{figure}[h!]
\centering
\includegraphics[angle=270,width=0.99\columnwidth]{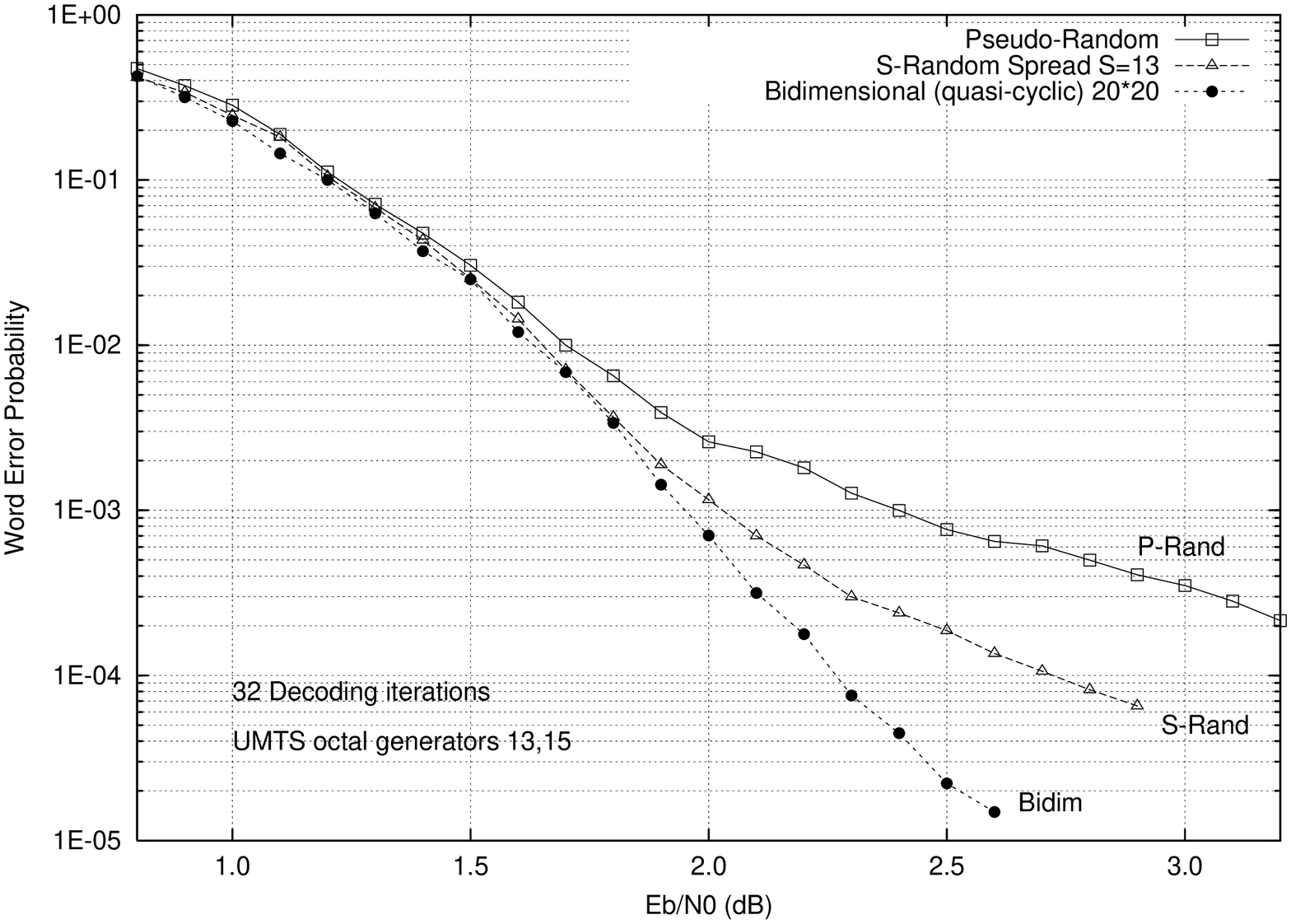}
\caption{Performance of rate 1/2 turbo code for different interleavers of size 400 bits. Octal generators (13,15), coding rate is raised from 1/3 to 1/2 by puncturing parity bits, 32 decoding iterations, additive white gaussian noise channel, binary phase shift keying modulation. For all points drawn above, at least 100 block errors and 500 bit errors have been measured during Monte Carlo simulation to estimate the word error probability.}
\label{fig:400}
\end{figure}

\begin{figure}[h!]
\centering
\includegraphics[angle=270,width=0.99\columnwidth]{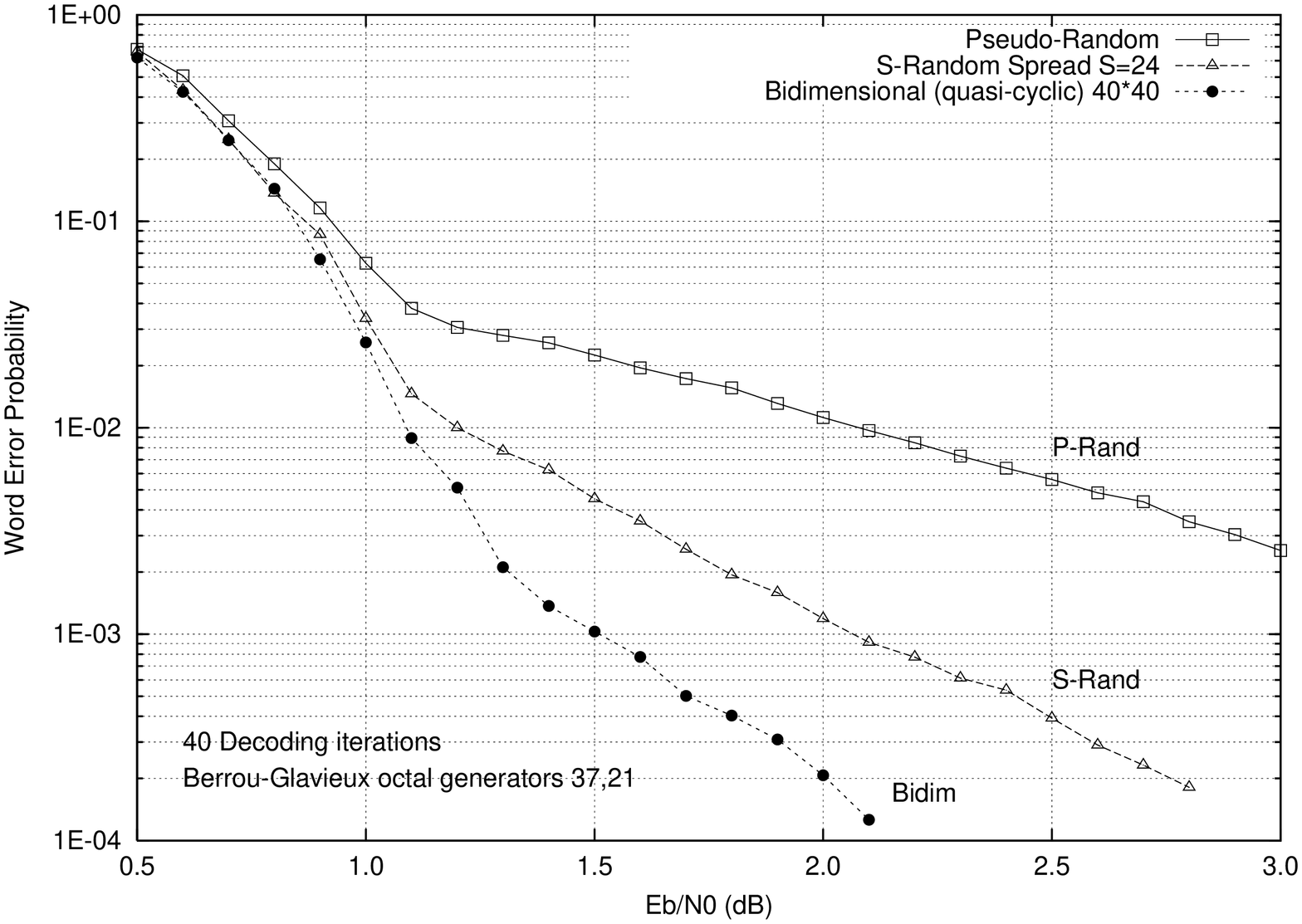}
\caption{Performance of rate 1/2 turbo code for different interleavers of size 1600 bits. Octal generators (37,21), coding rate is raised from 1/3 to 1/2 by puncturing parity bits, 40 decoding iterations, additive white gaussian noise channel, binary phase shift keying modulation. For all points drawn above, at least 100 block errors and 500 bit errors have been measured during Monte Carlo simulation to estimate the word error probability.}
\label{fig:1600}
\end{figure}

Word error rate versus signal-to-noise ratio 
results are depicted in Figures~\ref{fig:400} and \ref{fig:1600}.
In the first example, Fig. \ref{fig:400} illustrates the performance
of a rate 1/2 turbo code with an 8-state recursive systematic convolutional 
constituent $(13, 15)_8$. These octal generators have been adopted
in the European third generation mobile radio standard UMTS \cite{UMTS}.
The interleaver size is $N=400$, $n_1=n_2=20$, and the exact permutation is given
in Table \ref{Table400}. As shown in Fig. \ref{fig:400}, the bi-dimensional
interleaver clearly outperforms the spread interleaver. The increase
in minimum distance can also be validated numerically by measuring
the turbo code minimum distance 
using the algorithm proposed by Garello et.al. \cite{Garello2001}.

In the second example, the rate 1/2 turbo code has 
a recursive systematic convolutional constituent $(37,21)_8$, the octal generators 
proposed in the original turbo code by Berrou et.al. \cite{Berrou1993}.
The interleaver size is $N=1600$, $n_1=n_2=40$,  and the exact permutation is given
in Table \ref{Table1600}. As shown in Fig. \ref{fig:1600}, 
we get a significant improvement over the $S$-random interleaver.


\begin{table}
{\small
\begin{center}
\begin{tabular}{|r|l|}
\hline
\multicolumn{2}{|c|}{ \textbf{Square Bi-dimensional Quasi-cyclic Interleaver of Size 400}}\\\hline \hline
$\sigma$  & \texttt{~2 10 ~0 ~9 ~1 ~8 ~4 13 ~7 14 ~3 11 ~6 12 17 ~5 15 16 18 19} \\ \hline
$X$       & \texttt{~6 ~2 12 ~0 ~5 19 ~3 ~1 ~4 17 10 18 ~9 ~8 ~7 11 15 14 13 16} \\ \hline
\end{tabular}
\end{center}
}
\caption{Bi-dimensional interleaver of size $400=20 \times 20$. The first row defines the column permutation $\sigma$
and the second row defines the column cyclic shift $X$. This square interleaver is used in conjunction with RSC(13,15)
in Fig. \ref{fig:400}.}
\label{Table400}
\end{table}

\begin{table}
{\small
\begin{center}
\begin{tabular}{|r|l|}
\hline
\multicolumn{2}{|c|}{ \textbf{Square Bi-dimensional Quasi-cyclic Interleaver of Size 1600}}\\\hline \hline
$\sigma$  & \texttt{~1 15 17 18 25 39 33 29 19 ~4 ~0 37 14 20 27 ~9 22 31 10 28}\\ \hline
          & \texttt{30 36 23 35 ~7 16 ~6 ~2 13 26 ~3 34 32 21 11 ~8 ~5 38 12 24}\\ \hline
$X$       & \texttt{29 30 21 10 39 11 26 ~4 28 15 22 25 31 ~3 34 23 18 17 32 27} \\ \hline
          & \texttt{~0 ~9 ~1 19 24 36 ~2 37 ~6 35 14 33 20 13 ~8 12 ~5 16 38 ~7} \\ \hline
\end{tabular}
\end{center}
}
\caption{Bi-dimensional interleaver of size $1600=40 \times 40$. The first two rows define the column permutation $\sigma$ and the last two rows define the column cyclic shift $X$. This interleaver is used in conjunction with RSC(37,21)
in Fig. \ref{fig:1600}.}
\label{Table1600}
\end{table}

\pagebreak
\vspace*{2.2cm}

\pagebreak
\vspace*{2.2cm}

\begin{thebibliography}{99}

\bibitem{Abbasfar2004} A. Abbasfar and K. Yao,
``Interleaver design for turbo codes by distance spectrum shaping,''
in {\em Proc. IEEE Wireless Comm. and Net. Conf. (WCNC'04)},
vol. 3, pp. 1616-1619, Atlanta, March 2004.

\bibitem{Andrews1998} K.S. Andrews, C. Heegard and D. Kozen,
``Interleaver design methods for turbo codes,''
in {\em Proc. IEEE Int. Symp. Information Theory (ISIT'98)},
pp. 420, Cambridge, MA, Aug. 1998.

\bibitem{Bazzi2003} L. Bazzi, M. Mahdian and D. Spielman,
``The minimum distance of turbo-like codes,''
Submitted to the IEEE Transactions on Information Theory, May 2003.
Downloadable from http://math.mit.edu/$\sim$spielman/Research/mindist.html

\bibitem{Benedetto1996} S. Benedetto and G. Montorsi,
``Design of parallel concatenated convolutional codes,''
{\em IEEE Trans. on Communications}, vol. 44, no. 5, pp. 591-600, May 1996.

\bibitem{Berrou1993} C. Berrou, A. Glavieux, and P. Thitimajshima,
``Near Shannon limit error-correcting coding and decoding : turbo-codes,''
{\em Proceedings of ICC'93}, Geneva, pp. 1064-1070, May 1993.

\bibitem{Berrou1996} C. Berrou and A. Glavieux,
``Near optimum error correcting coding and decoding: Turbo-codes,''
{\em IEEE Trans. on Communications}, vol. 44, pp. 1261-1271, Oct. 1996.

\bibitem{Blahut2003} R.E. Blahut, Algebraic Codes for Data Transmission,
Cambridge University Press, 2003.

\bibitem{Breiling2001} M. Breiling,
``A Logarithmic Upper Bound on the Minimum Distance of Turbo Codes,''
Submitted to the IEEE Transactions on Information Theory, April 2001.
Downloadable from http://www.lnt.de/$\sim$breiling/Research/Publications/

\bibitem{Bravo2003} C.J. Corrada Bravo and P.V. Kumar,
``Permutation polynomials for interleavers in turbo codes,''
in {\em Proc. IEEE Int. Symp. Information Theory (ISIT'03)},
pp. 318, Yokohama, Japan, Jul. 2003.

\bibitem{Crozier1999} S.N. Crozier, J. Lodge,  P. Guinand, and A. Hunt,
``Performance of turbo codes with relative prime and golden interleaving strategies,''
in {\em Proc. of the 6th Int. Mobile Satellite Conf. (IMSC'99)},
Ottawa, Ontario, pp. 268-275, June 1999.

\bibitem{Crozier2000} S.N. Crozier,
``New high-spread high-distance interleavers for turbo-codes,''
in {\em Proc. of the 20th Biennial Symp. on Communications},
Queen's University, Kingston, Ontario, Canada, pp. 3-7, May 2000.

\bibitem{Crozier2001} S.N. Crozier and P. Guinand,
``High-performance low-memory interleaver banks for turbo-codes,''
in {\em Proc. of the 54th IEEE Veh. Tech. Conf. (VTC'01)},
Atlantic City, NJ, pp. 2394-2398, Oct. 2001.

\bibitem{Divsalar1995} D. Divsalar, F. Pollara,
``Turbo codes for PCS applications,''
in {\em Proc. IEEE Int. Conf. on Communications (ICC'95)},
vol. 1, pp. 54-59, Seattle, June 1995.

\bibitem{UMTS} European Telecommunications Standards Institute, 
``Universal mobile telecommunication system (UMTS); multiplexing and channel coding (FDD),''
3GPP TS 25.212, version 6.2.0, Release 6, 2004.

\bibitem{Gallager1963} R.G. Gallager, Low-density parity-check codes, MIT Press, 1963.

\bibitem{Garello2001} R. Garello, P. Pierleoni and S. Benedetto,
``Computing the free distance of turbo codes and serially concatenated codes with interleavers: Algorithms and applications,''
{\em IEEE Journal on Sel. Areas in Communications},
vol. 19, no. 5, pp. 800-812, May 2001.

\bibitem{Hokfelt1999} J. Hokfelt, O. Edfors and T. Maseng,
``Interleaver design for turbo codes based on the performance of iterative decoding,''
in {\em Proc. IEEE Int. Conf. on Communications (ICC'99)},
vol. 1, pp. 93-97, Vancouver, June 1999.

\bibitem{Kahale1997} N. Kahale and R. \"Urbanke,
``On the minimum distance of parallel and serially concatenated codes,''
1997. Downloadable from http://lthcwww.epfl.ch/publications/

\bibitem{Lin2004} S. Lin and D.J. Costello, 
Error Control Coding: Fundamentals and Applications, Prentice Hall, 2004.

\bibitem{Perez1996} L.C. Perez, J. Seghers, D.J. Costello, 
``A distance spectrum interpretation of turbo codes,''
{\em IEEE Trans. on Inf. Theory}, vol. 42, no. 6, pp. 1698-1709, Nov. 1996.

\bibitem{Peterson1972} W.W. Peterson and E.J. Weldon, Jr.,
Error-correcting codes, The MIT Press, 2nd edition, 1972.

\bibitem{Pothier2000} O. Pothier,
``Compound codes based on graphs and their iterative decoding,''
{\em PhD thesis in communications and electronics, ENST},
Paris, Jan. 2000. Downloadable from http://www.comelec.enst.fr/$\sim$boutros/coding

\bibitem{Ramsey1970} J.L. Ramsey,
``Realization of optimum interleavers,''
{\em IEEE Trans. Inform. Theory},
vol. IT-16, no. 3, May 1970.

\bibitem{Sadjadpour2001} H.R. Sadjadpour, N.J.A. Sloane, M. Salehi and G. Nebe,
``Interleaver design for turbo codes,''
{\em IEEE Journal on Sel. Areas in Communications},
vol. 19, no. 5, pp. 831-837, May 2001.

\bibitem{Tanner1999} R. M. Tanner,
``On Quasi-Cyclic Repeat-Accumulate Codes,''
Proc. of the 37th Annual Allerton Conference on Communication, 
Control, and Computing, Monticello, Illinois, Sept. 1999.

\bibitem{Tanner2000} R. M. Tanner,
``Toward an algebraic theory for turbo codes,''
in {\em Proc. of the 2nd International Symposium on Turbo Codes and Related Topics},
Brest, France, Sept. 2000.

\bibitem{Truhachev2001} D. V. Truhachev, M. Lentmaier and K. S. Zigangirov,
``Some results concerning design and decoding of turbo codes,''
Problemy Peredachi Informatsii, vol. 37, No. 3, 2001, pp. 190--205.

\bibitem{Vontobel2002} P.O. Vontobel,
``On the construction of turbo code interleavers based on graphs with large girth,''
in {\em Proc. IEEE Int. Conf. on Communications (ICC'02)},
vol. 3, pp. 1408-1412, New York, May 2002.

\bibitem{Yu2002} J. Yu, M.-L. Boucheret and R. Vallet,
``Design of turbo codes interleaver by loop distributions,''
in {\em Proc. IEEE Int. Symp. Information Theory (ISIT'02)},
pp. 54, Lausanne, Jul. 2002.

\end{thebibliography}
\end{document}